**DESCRIBING DIFFUSION, REACTION AND CONVECTION IN POROUS MEDIUM**

P.C.T. D´Ajello, L. Lauck and G.L. Nunes

Universidade Federal de Santa Catarina

Departamento de Física/ CFM

Brazil

P.O. Box 476 – CEP 88040-900

Fax 55 (48) 3721 9946

e-mail: pcesar@fisica.ufsc.br

**ABSTRACT**

In this paper we present a mathematical model for the electrochemical deposition aimed at the production of inverse opals. The real system consists of an arrangement of sub micrometer spheres, through which the species in an electrolytic medium diffuses until they react to the electrode surface and become part thereof. Our model consists in formulating convenient boundary conditions for the transport equation, that somewhat resembles the real system but is nevertheless simple enough to be solved, and then solve it. Similar approach was taken by Nicholson [1, 2], except that, to avoid the difficulties regarding the boundary conditions, he considered none whatsoever, and proposed a modified diffusion coefficient for the porous medium instead. Apropos, our model, with moving boundary condition pertain to the class of problems know as "The Stefan problem" [3].

## 1. INTRODUCTION

The porous system that we intend to model consists not only of such systems as shown in the work of Sapoletova and others [4,5,6], that show geometrical symmetries, but also irregular system as for instance biological diffusion processes. We consider that the one common aspect of all such systems is the specimen fluxes through well defined regions and the exchange of specimen among these regions and their neighbors. We model such regions simply as a right regular cylinder with periodical boundary as is shown in the Fig. 1. To mimic the interchange of particles we use a random process as described in the next section.

## 2. THE MODEL

We look for an expression for the current produced by the reduction and deposition of ions at the electrode surface. The electrode surface lies at the bottom of a cylindrical cavity. The cylindrical cavity is filled by an electrolytic solution that contains the ionic species, homogeneously diluted at the initial time.

In real systems the cross-sectional area of a diffusion path has a complex shape, dictated by the form of the spheres. For analytical tractability the model, however, will ignore those details and focus solely on the periodicity of the cross sectional area exhibited along the $z$ direction, assuming for it a cylindrical symmetry, with no angular dependence in the horizontal plane. The internal cross sectional radius of the vessel will vary, in a continuous and periodic fashion, from a minimum value $R_{\min}$ up to $R$, which is the external radius of the cylinder, as shown in Fig. 1(d). The radius depends on $z$ according the following relation,

$$R(z) = R\left[1 - \frac{1}{4}sin^2\left(\frac{\pi}{2R}z\right)\right].\tag{1}$$

The ionic concentration inside the finite cylindrical region obeys the balance equation:

$$\frac{\partial C}{\partial t} = D\left[\frac{\partial^2 C}{\partial r^2} + \frac{1}{r}\frac{\partial C}{\partial r} + \frac{\partial^2 C}{\partial z^2}\right] - \vec{v}\cdot\vec{\nabla}C,\tag{2}$$

where, $C = C(r, z, t)$ is the species concentration, which do not depend on the angular variable due to our assumption of rotational symmetry. In Eq. (2), $D$ is the diffusion constant and $\vec{v}$ the convective velocity, defined in the sense given by the Darcy's law [7], that is,

$$\vec{v} = v_c \hat{k}. \tag{3}$$

The model is completed by the following initial and boundary conditions:

$$C(r, z, 0) = c_b, \qquad\qquad (0 \leq r \leq R, z \geq 0) \tag{4a}$$

$$C(r, 0, t) = (c_b - c_s)e^{-kt} + c_s, \qquad (0 \leq r \leq R, \forall\, t) \tag{4b}$$

$$C(r, h, t) = c_b, \qquad\qquad (0 \leq r \leq R, \forall\, t) \tag{4c}$$

$$\left.\frac{\partial c}{\partial r}\right|_{r=0} = 0, \qquad\qquad (\forall\, z, \forall\, t) \tag{4d}$$

$$\left.\frac{\partial c}{\partial r}\right|_{r=R(z)} = -\alpha c_b(1 - e^{-vt})g(z)\,. \qquad (\forall\, z, \forall\, t) \tag{4e}$$

In Eqs.(2-4), $v_c$ is the magnitude of the convective velocity, $c_b$ the constant and homogeneous concentration in the cylindrical cavity at the initial time, before the applied electric potential, which allows for reduction of ions on the electrode, is turned on. $c_s$ is the limit concentration of species on the electrode. $k$ is the reaction rate; it describes the rate at which the surface transfers electrons to reduce the ions. In fact it is a function of the electric potential difference between the electrode and the liquid medium [8-12], which is taking as a constant due the assumption of a fixed electric potential difference. Therefore $k$ is a factor that defines the reaction kinetics. Despite its simplicity, boundary condition (4b) plays a central role in our description because it defines the concentration drop at the electrode where the ions are reduced and withdrawn from the liquid medium. In cases where $v_c = 0$, if $k = 0$ there is no reaction and there are no concentration changes at the surface ($C = c_b$ at any time). When $k \neq 0$, the concentration at the electrode surface evolves to a stationary value $c_s$ (when $t \to \infty$). A concentration gradient is produced and ions migrate from the bulk of the solution toward the surface. Boundary condition (4c) guarantees that we are working with a cylinder whose length is at least equal to the stationary diffusion layer h, which defines a distance from the electrode surface beyond which the ion concentration is assumed to be constant and equal to the bulk concentration $c_b$. Boundary condition (4d) is a consequence of the rotational symmetry of our system. Finally, the time dependent boundary condition (4e) determines the evolution of the concentration of ions that flow through the lateral surface of the cavity ($R$ (z) is the outer cylinder radius), in its normal direction. This boundary condition also depends on $z$, that is, the flux of matter through the cavity's lateral surface can change along its length. This boundary condition contains the parameter $\alpha$ (lower than one), which quantifies the magnitude of the matter flux that crosses the cylindrical lateral surface. Its signal determines the direction of matter flow, inward for negative $\alpha$ or outward if $\alpha$ is positive. The time dependent expression contained inside the brackets is a mathematical requirement to guarantee consistency of the boundary and initial conditions. Thus, at $t = 0$, there is no flow and the system is characterized by a constant and homogeneous distribution of matter ($C(r, z, t) = c_b$). When the potential is switched on, the symmetry is broken off. Ions begin to react at the cylinder bottom and a gradient arises so that species flow towards the electrode surface. In addition, the flow across the lateral area of the cylinder obeys a transient rule, quantified by the magnitude of the rate constant $\nu$ that appears on the exponential argument of Eq. (4e). A physical reasoning to justify the time dependent term in boundary condition (4e) is that $\nu$ quantifies the time interval elapsed until the flow assumes a magnitude that allows for the fluctuation of matter on the border of the cylinders according the rule prescribed by $g(z)$ and $\alpha$ on the lateral surface of the cavity. Regardless of this interpretation, we wish to emphasize the relevance of boundary condition (5e) for our model. It is the condition that sustains the flexibility of the model, that is, its capability to reproduce different situations, according to the sign of $\alpha$ and the form of the function $g(z)$. Thus, good choices of $\alpha$ and $g(z)$ make the model more real.

## 3. SOLVING THE BALANCE EQUATION

To simplify Eq. (2) we perform a variable transformation:

$$C(r, z, t) = U(r, z, t) exp\left(\frac{v_c}{2D} z - \frac{v_c^2}{4D} t\right),$$ (5)

which reads like,

$$C(r, z, t) = U(r, z, t) e^{\beta z - \Omega t},$$ (6)

after defining: $\frac{v_c}{2D} = \beta$, and $\frac{v_c^2}{4D} = \Omega$. Expression (5) transforms Eq. (2) into another equation,

$$\frac{\partial}{\partial t} U(r, z, t) = D\left[\frac{1}{r}\frac{\partial}{\partial r}\left(r\frac{\partial}{\partial r}U\right) + \frac{\partial^2}{\partial z^2}U\right],$$ (7)

and also transforms the boundary and initial conditions, now

$$U(r, z, 0) = c_b e^{-\beta z},$$ (8a)

$$U(r, 0, t) = (c_b - c_s)e^{-kt + \Omega t} + c_s e^{\Omega t},$$ (8b)

$$U(r, h, t) = c_b e^{-\beta h + \Omega t},$$ (8c)

$$\left(\frac{\partial U}{\partial r}\right)_{r=0} = 0, \tag{8d}$$

$$\left(\frac{\partial U}{\partial r}\right)_{r=R(z)} = -\alpha c_b(1 - e^{-\nu t})g(z)e^{-\beta z + \Omega t}. \tag{8e}$$

Because we search for a convenient form to compare the theoretical data and the experimental ones we compute the charge current that crosses the reactive surface in its normal direction z Thus, by another transformation we suppress the dependence in variable $r$ defining,

$$u(z,t) = \int_0^{R(z)} 2\pi r\, U(r,z,t)dr = \int_0^{R(z)} 2\pi r e^{-\beta z + \Omega t}C(r,z,t)dr. \tag{9}$$

This transformation simplifies the mathematical description without a limitation in the general characteristics of the systems given by our approach. As we will see on what follow the current density is obtained directly from the expression

$$J(r,z,t) = -D\bar{z}F\frac{\partial}{\partial z}C(r,z,t), \tag{10}$$

and still connected to the knowledge of the function $u(z,t)$ that came from a simple equation.

In Eq. (10) $\bar{z}$ is the charge number, connected to the number of electrons withdrawn from the electrode to reduce the ionic species on its surface. $F$ is the Faraday constant.

Knowing the relation between $u(z,t)$ and the current density we multiply Eq. (7) by $2\pi r$ to integrate in the variable $r$, so

$$\int_0^{R(z)} 2\pi r\frac{\partial U}{\partial t}dr = D\int_0^{R(z)} 2\pi r\frac{1}{r}\frac{\partial}{\partial r}\left(r\frac{\partial U}{\partial r}\right)dr + D\int_0^{R(z)} 2\pi r\frac{\partial^2 U}{\partial z^2}dr,$$

i.e.

$$\frac{\partial}{\partial t}\left[\int_0^{R(z)} 2\pi r U(r,z,t)dr\right] = 2\pi D\left[r\frac{\partial U}{\partial r}\right]_{r=R(z)} - 2\pi D\left[r\frac{\partial U}{\partial r}\right]_{r=0} + D\int_0^{R(z)} 2\pi r\frac{\partial^2 U}{\partial z^2}dr,$$

Which, because Eqs.(9), Eq. (8d) and Eq.(8e), reads:

$$\frac{\partial u(z,t)}{\partial t} = -2\pi DR(z)\alpha c_b H(t)g(z)e^{-\beta z + \Omega t} + D\int_0^{R(z)} 2\pi r\frac{\partial^2 U}{\partial z^2}dr, \tag{11}$$

Where, for simplicity we used

$$1 - e^{-\nu t} = H(t). \tag{12}$$

Before defining the differential equation for $u(z,t)$ let us consider the last term on the right hand side in Eq. (11). We know that, to any function $f(r,z,t)$ $\frac{d}{dz}\int_0^{R(z)} f(r,z,t)dr = \int_0^{R(z)} \frac{\partial f}{\partial z}dr + \frac{dR(z)}{dz}f((R(z),z,t),$

Thus, making

$f(r,z,t) = 2\pi r \frac{\partial U(r,z,t)}{\partial z}$ ,

We get,

$\frac{d}{dz}\int_0^{R(z)} 2\pi r \frac{\partial U}{\partial z}dr = \int_0^{R(z)} 2\pi r \frac{\partial^2 U}{\partial z^2}dr + \frac{dR(z)}{dz}2\pi R(z)\frac{dU(R(z),z,t)}{dz}.$ \hfill (A)

In similar fashion, given a function $g(r,z,t)$ follows the equality,

$\frac{d}{dz}\int_0^{R(z)} g(r,z,t)dr = \int_0^{R(z)} \frac{\partial g}{\partial z}dr + \frac{dR(z)}{dz}g$ ,

and assuming $g(r,z,t) = 2\pi r U(r,z,t)$ we obtain,

$\frac{d}{dz}\int_0^{R(z)} 2\pi r U(r,z,t)dr = \int_0^{R(z)} 2\pi r \frac{\partial U}{\partial z}dr + \frac{dR(z)}{dz}2\pi R(z)U(R(z),z,t).$ \hfill (B)

Rearranging terms in (B) in order to substitute $\int_0^{R(z)} 2\pi r \frac{\partial U}{\partial z}dr$ in (A), we obtain,

$\int_0^{R(z)} 2\pi r \frac{\partial^2 U}{\partial z^2}dr =$

$\frac{d^2 u(z,t)}{dz^2} - 2\pi\left(\frac{dR(z)}{dz}\right)^2 U(R(z),t) - 2\pi R(z)\frac{d^2 R(z)}{dz^2}U(R(z),t) - 4\pi R(z)\frac{dR(z)}{dz}\frac{dU(R(z),t)}{dz}.$ \hfill (C)

Because $U(r,z,t)$ is proportional to the concentration in just one point of the space, whereas $u(z,t)$ is the sum of concentration found in each point in a plane located at height $z$, then $u(z,t) \gg U(r,z,t)$. So, in good approximation:

$\int_0^{R(z)} 2\pi r \frac{\partial^2 U}{\partial z^2}dr \cong \frac{\partial^2 u(z,t)}{\partial z^2}.$ \hfill (13)

This approach is equivalent to assert that the flux of matter inside the cylindrical cavity is basically defined by the planar concentration gradient, with the planes arranged perpendicular to the $z$ axis.

Now we can use Eq. (13) to rewrite Eq.(11) as,

$\frac{\partial u(z,t)}{\partial t} = -2\pi R(z)D\alpha c_b H(t)g(z)e^{-\beta z+\Omega t} + D\frac{\partial^2 u(z,t)}{\partial z^2},$ \hfill (14)

and using Eq.(9) to define the new boundary and initial conditions (from Eqs. (8)) we get:

$u(z,0) = \pi[R(z)]^2 c_b e^{-\beta t},$ \hfill (15a)

$$u(0,t) = \pi[R(z)]^2(c_b - c_s)e^{-kt+\Omega t} + \pi[R(z)]^2 c_s e^{\Omega t}, \tag{15b}$$

and

$$u(h,t) = \pi[R(z)]^2 c_b e^{-\beta h + \Omega t} \ . \tag{15c}$$

Again we identify Eq. (14) as diffusion between planes with a source term on the border of the cylindrical cavity.

Our original problem now is reduced to a simpler equation (14) but still has non homogeneous boundary and initial conditions, equations (15a-15c). Before going further, we introduce a more compact notation, so we write:

$$u_t - Du_{zz} = q(z,t), \tag{16}$$

$$u(0,t) = A(t), \tag{17a}$$

$$u(h,t) = B^{(1)}(t), \tag{17b}$$

$$u(z,0) = B^{(2)}(z). \tag{17c}$$

An additional transformation makes homogeneous the boundary conditions, allowing us to use what is called the method of variation of parameters [15]. We choose a continuous and differentiable function $K(z,t)$ defined by,

$$K(z,t) = \frac{z}{h}B^{(1)}(t) - \frac{z-h}{h}A(t), \tag{18}$$

such $u(z,t)$, can be written as

$$u(z,t) = v(z,t) + K(z,t), \tag{19}$$

and, $v(z,t)$ is also a continuous and differentiable function of its variables.

Through the application of Eq.(19) into differential equation (16) we get,

$$v_t(z,t) - Dv_{zz}(z,t) = Q(z,t), \tag{20}$$

where,

$$Q(z,t) = q(z,t) - \frac{z}{h}B_t^{(1)} + \frac{z-h}{h}A_t(t), \tag{21}$$

with the boundary and initial conditions,

$$v(0,t) = u(0,t) - K(0,t) = 0, \tag{22a}$$

$$v(h,t) = u(h,t) - K(h,t) = 0, \tag{22b}$$

$$v(z,0) = u(z,0) - K(h,0) = B^{(2)}(z,0) - \frac{z}{h}B^{(1)}(0) + \frac{z-h}{h}A(0) = f(z). \tag{22c}$$

Now we have a partial differential equation with homogeneous boundary conditions. The method of variation of parameters [13] assumes solutions in the form,

$$v(z,t) = \sum_{n=1}^{\infty} T_n(t)\varphi_n(z), \tag{23}$$

where, $\varphi_n(z)$ are the eigenfunctions of the related homogeneous problem, i.e.:

$$\varphi_n(z) = sin(\omega_n z), \tag{24}$$

and $\omega_n = \frac{n\pi}{h}$ is the correspondent eigenvalue. Because the $\varphi_n(z)$ eigenfunctions are orthogonal functions, multiplying Eq. (24) by $\varphi_m(z)$, followed by an integration on the $z$ variable, gives

$$T_n(t) = \frac{2}{h}\int_0^h v(z,t)\varphi_n(z)dz. \tag{25}$$

Taking the derivative of Eq. (25), and using the expression (20) we get,

$$T_n'(t) = \frac{2}{h}\int_0^h Dv_{zz}\varphi_n(z)dz + Q_n(t), \tag{26}$$

where,

$$Q_n(t) = \frac{2}{h}\int_0^h Q(z,t)\varphi_n(z)dz, \tag{27}$$

and $T_n'(t)$ means the first derivative with respect to time.

Using Green formula,

$$\int_0^h v_{zz}\varphi_n(z)dz = [v_z\varphi_n - v\varphi_n']_0^h + \int_0^h v\varphi_n''dz, \tag{28}$$

and remembering that $v(z,t)$, as well as $\varphi_n(z)$ fulfill homogeneous boundary conditions, given by

$$\varphi_n'' = -\omega_n^2\varphi_n, \tag{29}$$

we get

$$T_n'(t) = -D\omega_n^2 T_n(t) + Q_n(t), \tag{30}$$

where use was made of Eq. (25). The integration of Eq. (30) turns to be an easy task after a multiplication by $e^{D\omega_n^2 t}$, so,

$$T_n(0) = \frac{2}{h}\int_0^h f(z)\varphi_n(z)dz = c_n, \tag{31}$$

$$T_n(t) = e^{-D\omega_n^2 t} c_n + e^{-D\omega_n^2 t} \int_0^t Q_n(s) e^{D\omega_n^2 s} ds, \tag{32}$$

And as a consequence,

$$v(z,t) = \sum_{n=1}^{\infty} sin(\omega_n z) \left[ c_n e^{-D\omega_n^2 t} + e^{-D\omega_n^2 t} \int_0^t Q_n(s) e^{D\omega_n^2 s} ds \right]. \tag{33}$$

To recover $u(z,t)$ we must invoke Eq. (19), such

$$u(z,t) = v(z,t) + K(z,t),$$

with

$$Q_n(s) = \frac{2}{h} \int_0^h \left[ q(z,s) - \frac{z}{h} \frac{\partial}{\partial s} B^{(1)} + \frac{z-h}{h} \frac{\partial}{\partial s} A \right] sin(\omega_n z) dz, \tag{34}$$

and

$$c_n = \frac{2}{h} \int_0^h \left[ B^{(2)}(z,0) - \frac{z}{h} B^{(1)}(0) + \frac{z-h}{h} A(0) \right] sin(\omega_n z) dz. \tag{35}$$

Remembering that

$$K(z,t) = \frac{z}{h} B^{(1)}(t) - \frac{z-h}{h} A(t),$$

$$A(t) = \pi[R(z)]^2 (c_b - c_s) e^{-kt+\Omega t} + \pi[R(z)]^2 c_s e^{\Omega t},$$

$$B^{(1)}(t) = \pi[R(z)]^2 c_b e^{-\beta h+\Omega t},$$

$$B^{(2)}(z) = \pi[R(z)]^2 c_b e^{-\beta z},$$

and

$$q(z,t) = -2\pi D R(z) \alpha c_b (1 - e^{-\nu t}) e^{-\beta z+\Omega t} g(z),$$

we have a lot of integral to be performed in order to obtain a defined value to $Q_n(s)$ and also to $c_n$. To perform this task we assume $h = mR$, that is, $h$ is an integer ($m = even$) number times $R$, and also $hR \ll 1$, given the magnitude of these parameters. Finally, using the abbreviate form $\tilde{g}_n$ to represent

$$\tilde{g}_n = \int_0^h R(z) g(z) e^{-\beta z} sin(\omega_n z) dz, \tag{36}$$

We get

$$Q_n(s) = \frac{2}{h}\left[-2\pi D\alpha c_b(1-e^{-\nu s})e^{\Omega s}\tilde{g}_n + \frac{\pi}{h}c_b\Omega e^{-\beta h+\Omega s}\frac{(-1)^n h^2 R^2}{n\pi} - \frac{\pi}{4h}(-k+\Omega)(c_b-c_s)e^{-ks+\Omega s}\frac{R^2 h^2}{n\pi} - \frac{\pi}{4h}c_s\Omega e^{\Omega s}\frac{R^2 h^2}{n\pi}\right].$$

(37)

Also $c_n$ has an expression

$$c_n =$$
$$\frac{2}{h}\left[\pi c_b \int_0^h [R(z)]^2 e^{-\beta z}\sin(\omega_n z)\,dz + \right.$$
$$\left. \pi c_b(1-e^{-\beta h})\int_0^h \frac{z}{h}[R(z)]^2\sin(\omega_n z)\,dz - \pi c_b\int_0^h [R(z)]^2\sin(\omega_n z)\,dz\right].$$

(38)

We do not show the result for the integrations in $c_n$ because we will not use it here. Anyway, Eq. (19) can be write as,

$$u(z,t) =$$
$$\frac{z}{h}\pi[R(z)]^2 c_b e^{-\beta h+\Omega t} - \frac{(z-h)}{h}\left[\pi[R(z)]^2(c_b-c_s)e^{-kt+\Omega t} + \pi[R(z)]^2 c_s e^{\Omega t}\right] +$$
$$\sum_{n=1}^{\infty}\sin(\omega_n z)\left[c_n e^{-D\omega_n^2 t} + e^{-D\omega_n^2 t}\int_0^t e^{D\omega_n^2 t}\frac{2}{h}\left\{-2\pi D\alpha c_b(1-e^{-\nu s})e^{\Omega s}\tilde{g}_n + \frac{\pi}{h}c_b\Omega e^{-\beta h+\Omega s}\frac{(-1)^n h^2 R^2}{n\pi} - \right.\right.$$
$$\left.\left. \frac{\pi}{h}(-k+\Omega)(c_b-c_s)e^{-ks+\Omega s}\frac{R^2 h^2}{n\pi} - \frac{\pi}{4h}c_s\Omega e^{\Omega s}\frac{R^2 h^2}{n\pi}\right\}ds\right].$$

(39)

The integrals appearing in Eq. (39) can be executed in order to give us

$$u(z,t) =$$
$$\frac{z}{h}\pi[R(z)]^2 c_b e^{-\beta h+\Omega t} - \frac{(z-h)}{h}\left[\pi[R(z)]^2(c_b-c_s)e^{-kt+\Omega t} + \pi[R(z)]^2 c_s e^{\Omega t}\right] +$$
$$\sum_{n=1}^{\infty}\sin(\omega_n z)\left[c_n e^{-D\omega_n^2 t} + \frac{2}{h}c_b(-2\pi D\alpha\tilde{g}_n)\left\{\frac{\left[e^{\Omega t}-e^{-D\omega_n^2 t}\right]}{\Omega+D\omega_n^2} - \frac{\left[e^{(\Omega-\nu)t}-e^{-D\omega_n^2 t}\right]}{\Omega-\nu+D\omega_n^2}\right\} + \right.$$
$$\frac{2}{h}\Omega(-1)^n\frac{\pi R^2 c_b}{\omega_n}e^{-\beta h}\frac{\left[e^{\Omega t}-e^{-D\omega_n^2 t}\right]}{\Omega+D\omega_n^2} - \frac{2}{h}\frac{\pi R^2}{\omega_n}(-k+\Omega)(c_b-c_s)\frac{1}{\Omega-k+D\omega_n^2}\left[e^{(-k+\Omega)t}-e^{-D\omega_n^2 t}\right] -$$
$$\left. \frac{1}{2h}\Omega c_s\frac{\pi R^2}{\omega_n}\frac{\left[e^{\Omega t}-e^{-D\omega_n^2 t}\right]}{\Omega+D\omega_n^2}\right].$$

(40)

Now we show how the current is obtained from the knowledge of $u(z,t)$.

The aim of this work is to determine the electrical current flowing through the electrode-solution interface. The flow of matter is not necessarily connected to the flow of charges spent by the system in order to reduce the ions on the electrode surface, so the density of charges is given by

$$J(r,z,t) = -D\bar{z}F\frac{\partial C(r,z,t)}{\partial z},$$

(41)

were $\bar{z}$ is the charge number and $F$ the Faraday constant. Because the total current flowing through the very base of the cylindrical cavity is given by

$$I(z,t) = -D\bar{z}F \int_0^{R(z)} 2\pi r \frac{\partial C(r,z,t)}{\partial z} dr \ , \tag{42}$$

using Eq. (6) we have

$$I(z,t) = -D\bar{z}F \int_0^{R(z)} 2\pi r \frac{\partial}{\partial z}\left[U(r,z,t)e^{\beta z - \Omega t}\right] dr, \tag{43}$$

and then

$$I(z,t) = -D\bar{z}F \left[2\beta u(z,t) + \frac{\partial u(z,t)}{\partial z} - 2\pi R(z)\frac{dR(z)}{dz}U(r,z,t)\right]e^{\beta z - \Omega t}. \tag{44}$$

To obtain Eq. (44) we used the definition given by Eq. (9). On the electrode surface i.e. were $z = 0$ and were this current is defined, the gradient of $u(z,t)$ has a relevant and central magnitude whereas the magnitude of $U(r,z,t)$ is negligible face the magnitude of $u(z,t)$ (see Eq. (9)) then we disregard the last term on the right side of Eq. (44). Now we will consider a particular case of this problem, which fits best to the experimental situation, we refer to the case where there is no convective movement, i.e.: $v_c = 0$, which means $\beta = \Omega = 0$ (see definitions that are below the Eq. (6)). One of the consequences of having $\beta = \Omega = 0$ is verified in the value assumed by $c_n$, i.e.: $c_n = 0$ (see Eq. (38)). Moreover, if we assume $c_s = 0$ we obtain,

$$I(z,t) = -D\bar{z}F\frac{\partial u(z,t)}{\partial z} =$$

$$-D\bar{z}F\left[\frac{\pi}{h}[R(z)]^2 c_b - \frac{2Rz\pi}{h}c_b R(z)\, sin\left(\frac{\pi}{R}z\right) - \frac{\pi}{h}[R(z)]^2 c_b e^{-kt} + \frac{z}{h}\pi c_b 2RR(z)sin\left(\frac{\pi}{R}z\right) + \right.$$

$$\frac{2}{h}\pi R^2 c_b k \sum_{n=1}^{\infty} cos(\omega_n z)\left\{\frac{e^{-kt}-e^{-D\omega_n^2 t}}{D\omega_n^2 - k}\right\} - \frac{2}{h}c_b\, 2\pi D\alpha \sum_{n=1}^{\infty} \tilde{g}_n \omega_n cos\, (\omega_n z)\left\{\frac{1-e^{-D\omega_n^2 t}}{D\omega_n^2} - \frac{e^{-vt}-e^{-D\omega_n^2 t}}{D\omega_n^2 - v}\right\}\left.\right] \ .$$

$$\tag{45}$$

To specify the current on the electrode it is enough to assume $z = 0$ in Eq. (45).

**CONCLUSIONS**

In this paper we demonstrated that an ordered porous system can be represented by a set of cylindrical vessels with permeable walls. A set of boundary and initial conditions for diffusion and reaction in a cylindrical vessel results in an analytical expression for the current transient measured at the bottom of the cylinder that is completely defined by a group of parameters that control the intensity of the ionic flux through the walls and the rate of ion consumption at the bottom of the cylinder. The introduction of a periodic corrugation in the cylindrical along with the assumption of a dissipative process that follows the random fluctuations in the path's line was essential to reproduce the current minima observed in experiments.

**FIGURE CAPTIONS**

**Figure 1**: (a) Top view of a colloidal crystal formed by four layers of spheres self-ordered on a flat substrate in a fcc structure. The star at the center indicates one pore through which an impinging ion may enter the porous structure. The write arrows depict one possible diffusion path towards the flat substrate. (b) Schematic cross section of the colloidal crystal. Black arrows indicate the diffusion path that was singled out in (a). (c) That particular diffusion path is now modeled as a staircased cylinder. The whole porous structure can be seen as a periodic replication of aligned twisted vessels. At each inflection point the vessels may exchange particles with neighboring ones. (d) In the last simplification step, the twisted vessels transform into straight cylinders with corrugated walls. The function g(z) is sketched on the right, indicating the points of maximum and minimum flux among vessels.

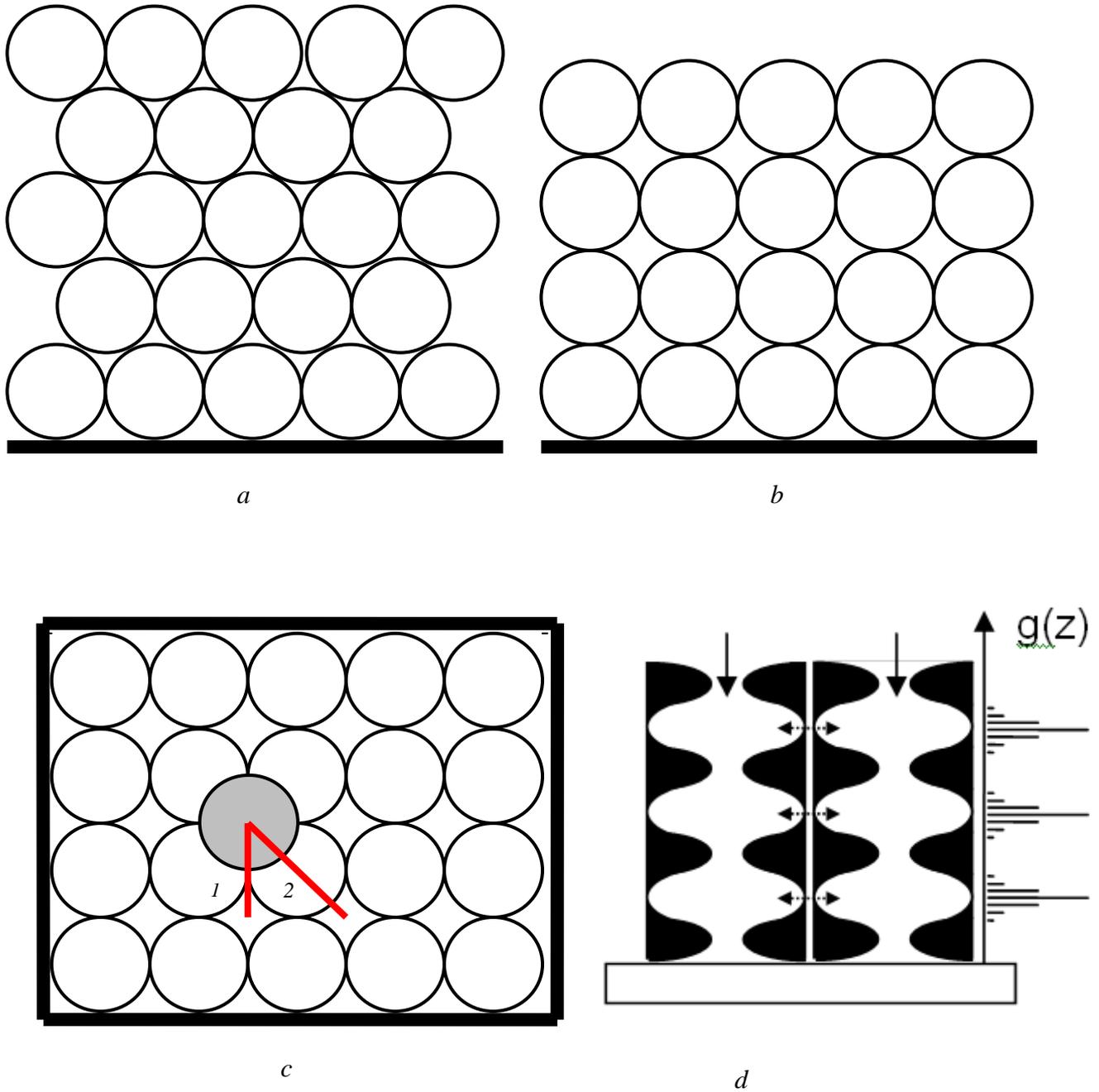

*Figure: 1*